\documentclass[final,authoryear,12pt]{elsarticle}

\usepackage{amssymb}
\usepackage{array,multirow}
\usepackage{amsmath}
\usepackage{color}
\usepackage{ulem}
\usepackage{soul}
\usepackage{xcolor}
\usepackage{booktabs}

\usepackage{lineno}
\journal{Earth and Planetary Science Letters}

\begin{document}

\begin{frontmatter}



\title{Single particle triboelectrification of Titan sand analogs}


\author[1]{Xinting Yu\corref{cor1}}
\ead{xyu33@jhu.edu}

\author[1]{Sarah M. H\"orst}

\author[1]{Chao He}

\author[2]{Patricia McGuiggan}

\cortext[cor1]{Corresponding author}
\address[1]{Department of Earth and Planetary Sciences, Johns Hopkins University, 3400 N. Charles Street, Baltimore, Maryland 21218, USA}
\address[2]{Department of Materials Science and Engineering, Johns Hopkins University, 3400 N. Charles Street, Baltimore, Maryland 21218, USA}
\begin{abstract}
Sand electrification is important for aeolian sediment transportation on terrestrial bodies with silicate sand as the main sediment composition. However, it has not been thoroughly studied for icy bodies such as Titan with organic sand as the main dune-forming material. We used the colloidal probe atomic force microscopy (AFM) technique to study triboelectric charging processes using Titan and Earth sand analogs. We found that it is easy to generate triboelectric charges between naphthalene (a simple aromatic hydrocarbon), polystyrene (an aromatic hydrocarbon polymer), and borosilicate glass (Earth silicate sand analog). Strong electrostatic forces can be measured after contact and/or tribocharging. In contrast, tholin, a complex organic material, does not generate any detectable electrostatic forces with contact or tribocharging within the detection limit of the instrument. If Titan sand behaves more like tholin, this indicates that the tribocharging capacity of Titan sand is much weaker than Earth silicate sand and much less than previously measured by M\'endez-Harper et al., (2017), where only simple organics were used for Titan sand analogs. Thus, triboelectrification may not contribute to increasing interparticle forces between sand particles on Titan as much as on Earth. Interparticle forces generated by other electrostatic processes or other interparticle forces such as van der Waals and capillary cohesion forces could be the dominant interparticle forces that govern Titan sand formation and sediment transportation on the surface. Titan sand is also unlikely to produce large electrical discharge through tribocharging to affect future missions to Titan's surface.
\end{abstract}





\end{frontmatter}

\begin{itemize}
\item Research highlight 1: Simple organics and polymers charge strongly through contact and tribocharging.
\item Research highlight 2: Titan aerosol analog, or ``tholin", produces minimal triboelectric charges.
\item Research highlight 3: The charging capacity of Titan sand may be much weaker than previously measured.
\item Research highlight 4: Titan sand may be unlikely to produce large electrical discharge by tribocharging.
\end{itemize}

\section{Introduction}
Aeolian landforms have been observed on various planetary bodies with solid surfaces, despite the diverse environmental conditions such as surface temperature (from $<$40 K for outer solar system bodies such as Pluto and Triton to $\sim$740 K for Venus) and surface atmospheric pressure (from $<$10$^{-5}$ bar to $\sim$90 bar) on these bodies. Other than environmental conditions, the transporting materials on these bodies are different as well. For the inner solar system bodies, Venus, Mars, and Earth, the main transporting materials are silicate sand (Greeley \& Iversen, 1985), while it is mainly ices or organics transporting on the surfaces of outer solar system objects including Titan (Lorenz et al., 2006), Pluto (Telfer et al., 2018), and Triton (Smith et al., 1989).

On Titan, the dune particles are interpreted to be made of mainly organics or organic coated ices (e.g., McCord et al., 2006; Soderblom et al., 2007; Clark et al., 2010). There are several possible candidates that could make up Titan's sand particles (see e.g., Barnes et al., 2015; Yu et al., 2018), including complex tholin-like organics, evaporites (dissolved part of the complex organics in Titan's liquid hydrocarbon lakes and seas), and water ice that makes up the bulk of Titan's crust.

One of the important parameters required to characterize aeolian sand transportation is the threshold wind speed, the minimum wind speed needed to saltate sand particles. The threshold wind speed can be derived by a force balance between gravity, wind drag and lift forces, and the interparticle forces (Shao \& Lu, 2000). Gravity, wind drag and lift forces are mostly affected by the environmental conditions such as gravity and atmospheric density (they are also affected by particle density, see Yu et al., 2017a), while the interparticle forces are not only affected by environmental conditions, but also intrinsic material properties such as surface energy, surface chemistry and functional groups, surface roughness, etc. Yu et al., (2017b) has recently measured the surface energy of one possible Titan sand analog, "tholin", to be around 70 mN/m.

The interparticle forces for wind-blown sand generally include van der Waals forces, capillary forces due to condensed liquid, and electrostatic forces. Yu et al. (2017b) used atomic force microscopy (AFM) to directly measure the interparticle cohesion between Titan aerosol analog tholin particles, and they found that the van der Waals cohesion between tholin particles is larger than between silicate sand and analog materials used in the Titan Wind Tunnel (TWT), where experiments are conducted in ambient air. The above experiments mainly focused on measuring interparticle forces including van der Waals forces and capillary forces (when exposed in humid air), while the contributions from the electrostatic forces are small. In large-scale aeolian dune settings, the saltating grains are constantly colliding and rubbing together, which could induce electrostatic forces by contact charging and friction induced tribocharging. 

Contact charging, or contact electrification, is the electrical charge that developed between two materials (identical or different) during their contact and separation. Tribocharging also includes rubbing (friction) of the two materials during contact, thus increasing the applied force and frictional energy, and generally produces higher charge density. Both charging processes depend on the physical and chemical properties of the charging materials, environmental conditions (e.g., pressure, temperature and relative humidity (RH)), and the nature of the contact between the surfaces of the charging materials such as contacting geometry and surface roughness (Lacks \& Mohan-Sankaran, 2011).

Electrostatic forces act over a much longer range compared to van der Waals forces, and thus could affect both the initiation of sediment entrainment (e.g., the threshold wind speed) and particle trajectories after the sediments are lifted off the surface. Furthermore, charged particles injected high in the air could form strong electric fields with the particles near the surface or the surface itself, and would further alter the trajectories and charging properties of the saltating particles (Schmidt et al., 1998; Zheng et al., 2003; Kok and Renno, 2008). Chemistry (tribochemistry) could also be driven by the charging processes, where the mechanical energy could be turned into chemical energy during a local discharge, and could alter the chemical properties of the charged grains (Thomas and Beauchamp, 2014; Wu et al., 2018).

Sand electrification has proven to be an important process for terrestrial wind-blown sand transport. Electric fields and charges on sand grains have been characterized both in the field (e.g., Schmidt et al., 1998; Zhang et al., 2004) and in wind tunnel simulation experiments (e.g., Greeley \& Leach, 1978, Zheng et al., 2003; Zhang et al., 2004). The near surface electric fields can be up to tens to hundreds kV/m in a saltating sand cloud, pointing upward, which indicates a negatively charged surface and positively charged particles (Zheng et al., 2013). It is consistent with particle charge distributions measured in the wind tunnel experiments, where larger particles and the surface tend to charge positively, and smaller particles tend to charge negatively. Sand and dust electrification may also be important for aeolian transport and geologic features on Mars (Krauss et al., 2003; Anderson et al., 2009) and regolith levitation on the Moon (Singer \& Walker, 1962; Sickafoose et al., 2001). The materials tested in the above studies were mainly terrestrial silicate sand (Earth quartz sand, lunar or Mars regolith simulants), while sand electrification studies that include analog materials for icy bodies are necessary to fully understand aeolian processes in the outer solar system such as Titan.

M\'endez-Harper et al. (2017) measured electrification of some organic molecules as Titan sand analogs by using a tumbler to charge the grains and then measure the charge distribution of the particles using a Faraday cage. They found that the organic molecules that they investigated have higher maximum charge densities in a dry atmosphere (RH$<$1\%) than silicate sands in a humid atmosphere (30\%$<$RH$<$40\%). However, the above experiments used only simple organics and polymers (naphthalene, biphenyl, and polystyrene) as Titan sand analogs. An experiment that uses the Titan aerosol analogs, tholin, which could be the precursor of Titan sand by coagulating into bigger sand-size grains (e.g., Yu et al., 2017b), is needed to fully characterize the whole spectrum of the charging abilities of Titan sand candidates.

The difficulty with using tholin is that this material is usually produced in low yields (e.g., He et al., 2017), prohibiting its use in bulk studies. In this study we used a novel colloidal probe AFM technique to directly charge and measure electrostatic forces between single particles so that we can accommodate the small material volume of tholin. This technique has been used to measure the cohesion forces between tholin particles and tholin-coated borosilicate glass beads (Yu et al., 2017b). In this study, in order to reduce the effect of local surface roughness and geometry, we used smooth spheres (both bare or coated with other materials) and flat smooth surfaces for our measurements. All the measurements were done under dry conditions (RH$<$1\%). Controlling the contacting geometry and environmental conditions can reduce contribution from these factors in electrification, while enhancing the differences from the physical and chemical nature of the materials themselves. The tumbler experiment (M\'endez-Harper et al., 2017) involves a large number of variables by using materials of various geometries and surface roughness, and charging the terrestrial and Titan analog materials under different humidity conditions, which could complicate the interpretation of results on the intrinsic charging capacity of different materials.

Furthermore, with the precise control of both the lateral and vertical movements of the AFM, we can accurately characterize the process of contact charging and tribocharging individually, while the tumbler experiment (M\'endez-Harper et al., 2017) could involve electrification by multiple processes, not only contact charging and tribocharging, but also fragmentation charging (exchange of bits of charged materials between contact surfaces). Performing the experiments on single particle scale could also ensure that charging occurs only between the same material of the interest, while eliminating the effect of charging between dissimilar materials. 

A more detailed description of the technique and our experimental setup can be found in sections 2.1 and 2.2. In section 3.1, we test the technique using two materials (mica and silicate glass) that are known to charge each other (e.g., McGuiggan, 2008). In section 3.1--3.3, we present the triboelectrification results for borosilicate glass, a simple aromatic hydrocarbon naphthalene, an aromatic hydrocarbon polymer polystyrene, and a complex organic material called tholin, respectively. We discuss the possible charging mechanisms for the tested materials in section 4.1, and the implications of our results on aeolian transport and lightnings on Titan in sections 4.2 and 4.3.

\section{Methods}

\subsection{AFM Colloidal Tip and Surface Preparation}

Tholin was produced using the Planetary HAZE Research (PHAZER) experimental system at Johns Hopkins University, with a 5\% $\mathrm{CH_4/N_2}$ cold gas mixture (around 100 K) in a glow plasma discharge chamber (see He et al., 2017). Tholin was deposited homogeneously on the acid-washed microspheres, colloidal AFM probes (AFM cantilevers with a 20 $\mu$m diameter glass sphere attached to the end of the cantilever), and on mica discs (10 mm diameter). The coated tholin film is very smooth (RMS roughness is $\sim$1 nm for 1$\mu$m$\times$1$\mu$m scan, measured by AFM) and has a thickness of approximately 1.3 $\mu$m.

Two sources of glass spheres were used: soda lime glass microspheres of diameter 50--100 $\mu$m (Polysciences, Inc.) and sQube (Nanoandmore) 20 $\mu$m diameter borosilicate glass spheres. The 50--100 $\mu$m spheres were supplied in powder form and needed to be attached to the AFM cantilever. To do this, the microspheres were twice washed with a dilute HCL solution ($\sim$1.05 M/L) and then rinsed with HPLC-grade water eight times in an ultrasonic cleaner (Lab Safety Supply). They were then dried in a 65 $^{\circ}$C oven (Lab Safety Supply Model No.32EZ28) overnight. After coating with tholin, a single microsphere was then attached to the AFM cantilever using the AFM motors to position the sphere at the end of the cantilever and epoxy resin to glue the sphere to the end of the cantilever. The resulting colloidal probe was left to sit overnight in ambient air (McGuiggan et al., 2011; Yu et al., 2017b) to allow epoxy to dry. The sQube colloidal probe AFM cantilevers with a 20 $\mu$m diameter glass sphere attached were also purchased. These cantilevers were directly coated with tholin. This procedure allowed a shorter exposure of tholin to ambient air. We use spheres of diameter between 20-100 $\mu$m, because the AFM cantilevers we used has a limited spring constant of around 40 N/m, and larger spheres could break the cantilever easily due to gravity.

We made the glass and the polystyrene colloidal AFM probes by gluing a bare acid-washed glass sphere and a polystyrene sphere (diameter range: 85-105 $\mu$m, Cospheric, Santa Barbara, CA) to AFM cantilevers. To make naphthalene-coated colloidal AFM probe, we dip-coated a glass colloidal AFM probe into a naphthalene-acetone solution and then dried for 30 minutes. The naphthalene surface was made by putting a few drops of naphthalene-acetone solution on a glass slide and then dried for 30 minutes.

\subsection{Charging and Electrostatic Force Measurements}
We performed all the charging and force measurements using a Bruker Dimension 3100 atomic force microscope, which has vertical resolution of 0.1 nm. The AFM was enclosed in a controlled RH environment equipped with a digital hygrometer (Dwyer Instrument, RH range: 0--100\% with $\mathrm{\pm}$2\% accuracy, temperature range: -30--85 $^{\circ}$C with 0.5 $^{\circ}$C accuracy). The RH in the system can vary between $<$1\% in a dry nitrogen environment, to about 40--50\% in ambient air. Before each charging experiment, we flowed dry nitrogen into the AFM system enclosed in a glove bag and kept the RH$<1$\% for at least two hours. The spring constants of the AFM cantilevers are around 40 N/m, and were calibrated by thermal tuning. The sensitivity of the AFM photodiode was calibrated by indenting a hard silicon surface.

We performed both contact charging and tribocharging between the AFM colloidal probes and various surfaces. The AFM setup is shown in Figure 1(a). The colloidal particles on the colloidal probes include soda lime glass spheres (50--100 $\mu$m), polystyrene spheres (85--105 $\mu$m, Cospheric), tholin-coated glass spheres (50--100 $\mu$m and 20 $\mu$m for colloidal probes from sQube), and naphthalene-coated glass spheres (50--100 $\mu$m). The planar surfaces we used include cleaved mica sheets, tholin-coated mica sheets, and naphthalene-coated glass slides. All of the charging experiments were conducted with RH$<$1\% in dry nitrogen to prevent charge leakage through water in the air and oxygen contamination of the samples. An insulating PET film was placed beneath the surfaces to make sure that the generated electrostatic charges are not conducted outside the system. The experiments we conducted are summarized in Table \ref{table:materials}. The charging method was adopted and modified from Bunker et al., (2007) and is described below.

For contact charging, we used the colloidal probe to repeatedly contact the substrate in the normal direction by taking multiple force-distance curves at a rate of 1 Hz without lateral scanning. For tribocharging, the colloidal probe was scanned over a 10 $\mu$m$\mathrm{\times}$10 $\mu$m area for 256 lines using contact mode imaging with a scan speed of 3 Hz. Contact mode was used to ensure that rubbing and friction between the colloidal probe and the surface is occuring during the imaging. Immediately after the scanning was completed, a series of force curves were collected between the colloidal probe and the substrate. An example force curve and the interactions between the colloidal probe and the surface are shown in Figure 1(b). Because of the dry environment (RH$<$1\%), no capillary forces should be involved in the jump-in and pull-off forces. The jump-in forces could include short-range van der Waals forces and long-range electrostatic forces if generated by charging. The pull-off forces are an indication of the total cohesion (between the same materials) or adhesion (between two different materials) forces.

We can also vary the scan speed and scanning area to test different tribocharging strengths. The scan speed can be tuned from 0.5 Hz to 10 Hz, which are equivalent to between 10 $\mu$m/s to 200 $\mu$m/s. We can also vary the scanning area from 10 $\mu$m$\mathrm{\times}$10 $\mu$m, to up to 50 $\mu$m$\mathrm{\times}$50 $\mu$m. After the charging experiments, the colloidal probes were imaged using scanning electron microscopy (SEM) to examine the surface conditions and to measure for exact size of the colloidal spheres.

The minimum charge (Q$_{\textrm{lim}}$) detectable by the AFM between two chemically identical particles with radius R$_1$ and R$_2$ can be estimated as follows. Assuming the net charges generated on each surface is +Q and -Q, the electrostatic interaction forces (F$_E$) can be written as (Bichoutskaia et al., 2010):
\begin{equation}
F_E=K\frac{(+Q)(-Q)}{r^2}-f(k,R_1,R_2,r),
\label{eq:force}
\end{equation}
where K=1/4$\pi \varepsilon_0$ $\approx$ 9$\times$10$^9$ V$\cdot$m/C, r is the distance between the two particles, and k is the dielectric constant of the material. Because the total net charge of the two surfaces has to be zero, the first term is always negative (attractive force). The second term describes the polarization effect and is always attractive (Bichoutskaia et al., 2010). Here we use the first term to calculate the detection limit, which provides an upper limit of the detectable minimum charge. 

The vertical resolution of the AFM is 0.1 nm, but to confirm that long-range force exists between the surfaces, not fluctuation of the baseline, a vertical range deviation to the baseline of at least $\sim$5 nm is needed. We can assume a typical spring constant of the AFM cantilever to be 40 N/m, thus the minimum detectable force by the AFM is 80 nN. The electrostatic interaction typically happens in a distance range between 0.1 nm and 100 nm. Using the first term of Equation \ref{eq:force}, the minimum charge detection limit Q$_{\textrm{lim}}$ is thus 1.3$\times$10$^{-18}$ to 1.3$\times$10$^{-15}$ C, which is lower than the detection limit of common charge amplifiers (e.g., Q$_{\textrm{lim}}\approx$ 10$^{-14}$ C, M\'endez-Harper et al., 2017). 

\begin{table}
\caption{Summary of materials used in this study. The electrical resistivity of tholin is from Khare \& Sagan (1980). The electrical resistivity of naphthalene is from Wikipedia. The electrical resistivity of polystyrene and glass are from the Engineering Toolbox.}
 \label{table:materials}
 \centering
 \begin{tabular}{c c c c c}\toprule
Colloidal Sphere & Sphere Electrical & Substrate Material & Contact  & Tribo- \\
Material&   Resistivity ($\Omega\cdot m$) & &   Charging &charging  \\
\midrule
\multirow{2}{*}{Coated tholin} & \multirow{2}{*}{10$^6$--10$^7$} & Tholin film & No & No\\
&& Tholin coated sphere & No & No\\
\hline
Coated napthalene  & 10$^{12}$ & Napthalene film & Yes & Yes\\
\hline
Polystyrene  & $\sim$10$^{14}$ & Polystyrene sphere & Yes & Yes\\
\hline
\multirow{2}{*}{Glass } &\multirow{2}{*}{10$^{10}$--10$^{11}$} & Fused silica slide & Yes & Yes\\
 & & Cleaved mica sheet & No & Yes\\
\bottomrule
 \end{tabular}
 \end{table}

\begin{figure}[h]
\centering
\includegraphics[width=25pc]{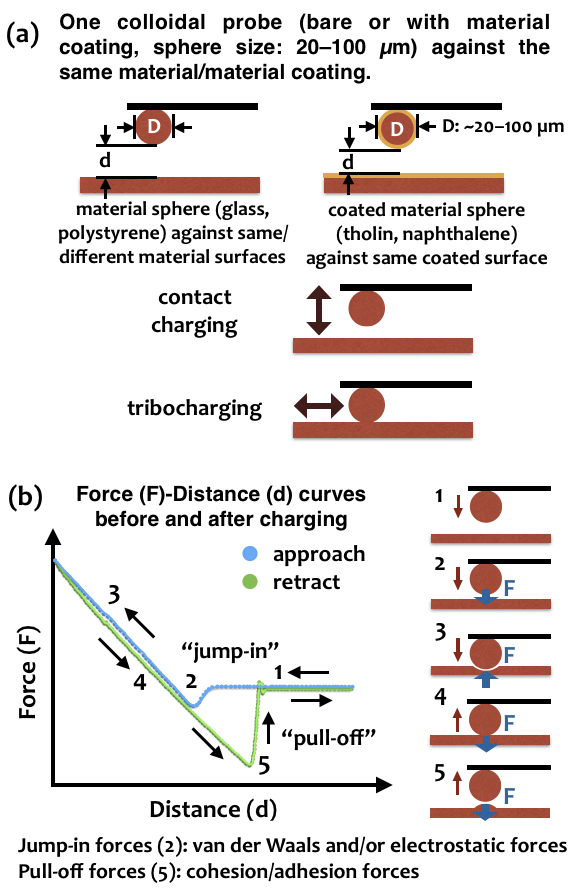}
\caption{(a) The AFM setup used for the charging and electrostatic force measurements. All the experiments were done under RH\textless1\%. Contact charging was done by repeated contacting the material particle to the material surface, while tribocharging was done by using the material particle to scan across the material surface. (b) Before and after each charging experiment, force-distance curves were taken using the AFM. Here is a typical force curve between a colloidal probe and a smooth surface, the different interaction phases are described in Yu et al., (2017b). Here we mostly focus on the jump-in forces before the particle contacts the surface and the pull-off forces when the particle was retracted from the surface. The jump-in forces in our study could include short-range van der Waals forces and long-range electrostatic forces, while the pull-off forces are equal to the adhesion/cohesion forces between the probe and the surface.}
\label{fig:1}
\end{figure}

\section{Results}

\subsection{Validation of methodology}
\begin{figure}[h]
\centering
\includegraphics[width=30pc]{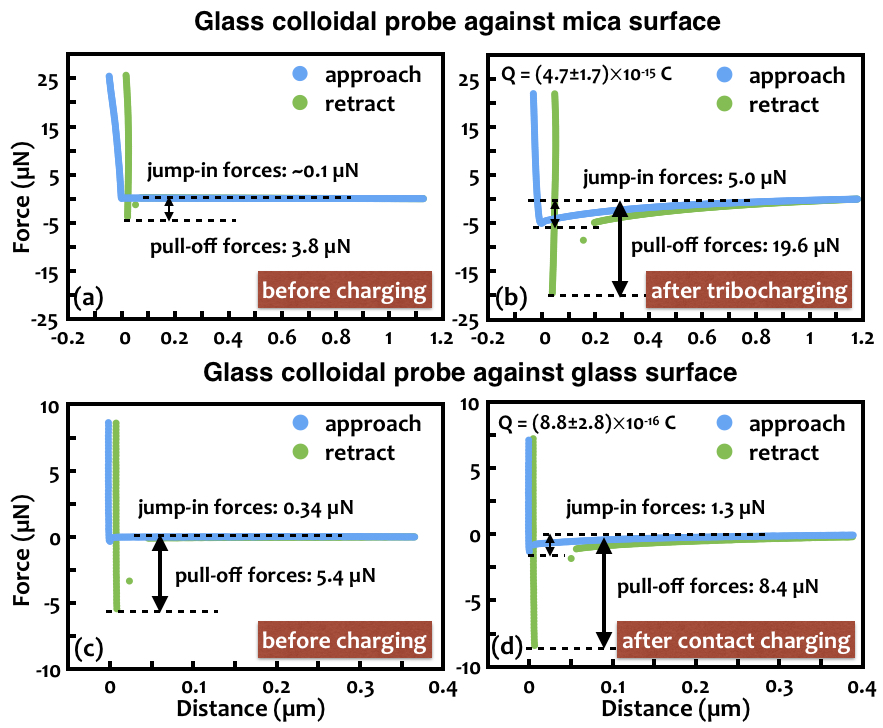}
\caption{Force-distance curves taken between a glass colloidal AFM probe and a cleaved mica surface before and after a series of charging events in a dry nitrogen atmosphere (RH$<$1\%). (a) Force curve before charging, it has very weak jump-in forces and some short-range pull-off forces. (b) Force curve after contact charging, which is similar to the force curve before charging. (c) The force curve immediately after tribocharging. The jump-in and pull-off forces after charging become both larger and in longer range than before charging. (d) The 30th force curve after tribocharging shows charge damping. The jump-in forces are not long range any more but the magnitude of the pull-off forces remains similar as just after tribocharging.}
\label{fig:4}
\end{figure}

To test our experimental setup, we performed charging experiments initially between two different materials that are known to charge each other, fused silica (soda lime glass sphere) and mica (e.g., McGuiggan, 2008) in a nitrogen environment (RH$<$1\%). We measured the force-distance curves before charging, after contact charging, and after tribocharging. We selected the last set of force curves before charging, and the first sets of force curves right after tribocharging and show them in Figure \ref{fig:4}. Before charging, the jump-in forces are very small, and the pull-off forces are short-range, around 3.8 $\mu$N (Figure \ref{fig:4}(a)). After contact charging, the jump-in forces remain small and short-range, which indicates that no long-range electrostatic interaction was generated from contact charging. The force curve immediately after tribocharging (Figure \ref{fig:4}(b)) shows that both the jump-in forces and pull-off forces become larger and longer-range (both forces act over a range of 0.8--1 $\mu$m). The jump-in forces are two orders of magnitudes larger than before charging, $\sim$5.0 $\mu$N, and the pull-off forces are about 5 times larger than before charging, $\sim$19.6 $\mu$N. It seems electrification can greatly enhance adhesion between materials, which is observed also by Bunker et al., (2007). Through the long-range jump-in and pull-off forces, we can calculate the amount of charge generated on the particle. Using the first term of Equation \ref{eq:force}, we can estimate the charge amount (Q) of (4.7 $\pm$ 1.7)$\times$10$^{-15}$ C. With SEM image, we are able to measure the size of the glass sphere and then calculate the charge to mass ratio (Q/m) to be (9.1 $\pm$ 3.4)$\times$10$^{-6}$ C/kg (glass sphere diameter 73 $\mu$m and density of sphere 2500 kg/m$^3$).

We also performed charging between two identical glass surfaces (a colloidal glass sphere and a glass surface) that are reported to charge in granular systems (Forward et al., 2009). As shown in Figure \ref{fig:4}(c), both the jump-in forces and the pull-off forces are short-range before charging. After contact charging, both the jump-in forces and the pull-off forces become longer in range and larger in magnitude. The jump-in forces increase from 0.34 $\mu$N to 1.3 $\mu$N (4 times increase), and the pull-off forces increase from 5.4 $\mu$N to 8.4 $\mu$N ($\sim$1.5 times increase), both forces act over 0.3 $\mu$m after contact charging. The result is similar after tribocharging. We notice that the extent of charging between chemically identical materials is weaker than between different types of materials. The increases in both force magnitudes and the range of forces between glass surfaces are less than between glass-mica surfaces.  The calculated charge amount (Q = (8.8 $\pm$ 2.8)$\times$10$^{-16}$ C) and the charge to mass ratio (Q/m = (9.3 $\pm$ 2.9)$\times$10$^{-7}$ C/kg, glass sphere diameter 90 $\mu$m and density of sphere 2500 kg/m$^3$, see also Figure \ref{fig:5}) are also an order of magnitude smaller between the two glass surfaces compared to between glass and mica surface. The charge to mass ratio between two glass surfaces is within the measured charge to mass ratio range of silicate sand measured by M\'endez-Harper et al., (2017).

\subsection{Charging between simple organics}

\begin{figure}[h]
\centering
\includegraphics[width=30pc]{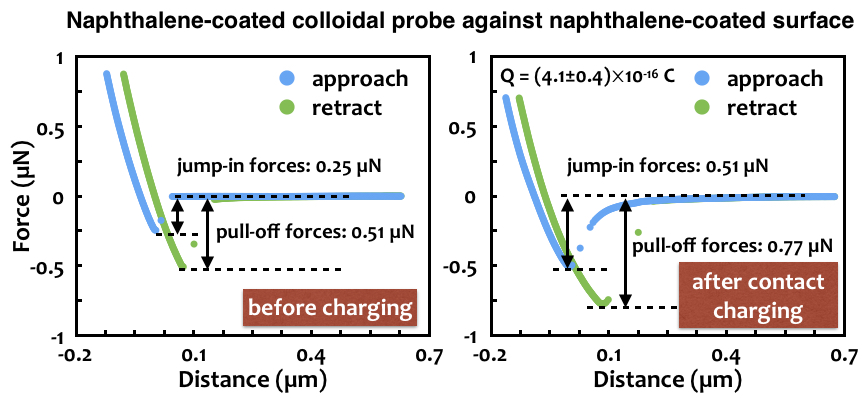}
\caption{Force-distance curves taken between a naphthalene-coated colloidal AFM probe and naphthalene-coated surface, before charging and after contact charging. The jump-in and pull-off forces after charging become both larger and in longer range than before charging.}
\label{fig:2}
\end{figure}

We performed charging experiments between the simple PAH naphthalene and the hydrocarbon polymer polystyrene in a dry nitrogen environment (RH$<$1\%). Naphthalene is a possible constituent of Titan's atmosphere (Waite et al., 2007), and both naphthalene and polystyrene were used by M\'endez-Harper et al., (2017) as the Titan sand analog materials. Figure \ref{fig:2} shows the force curves taken between a coated naphthalene microsphere and a coated naphthalene surface, right before and after contact charging. We were able to generate strong long-range electrostatic attraction by continuously contacting the two surfaces. As shown by the approach curves, after charging, the jump-in forces became two times bigger from 0.25 $\mu$N to 0.51 $\mu$N. The jump-in forces also act for about an order of magnitude longer range ($>$0.3 $\mu$m) compared to before charging (0.04 $\mu$m). Note that the pull-off forces also become larger (from 0.51 $\mu$m to 0.77 $\mu$N) and act longer range (from 0.08 $\mu$m to 0.13 $\mu$m) than before charging, which suggests electrification also enhances cohesion between materials. The calculated charge amount is (4.1 $\pm$ 0.4)$\times$10$^{-16}$ C and the charge to mass ratio is (1.4 $\pm$ 0.1)$\times$10$^{-6}$ C/kg for the force curve after contact charging (sphere diameter 60 $\mu$m and density of sphere 2500 kg/m$^3$, Figure \ref{fig:5}), which is in within the measured charge to mass ratio range by M\'endez-Harper et al., (2017). Tribocharging also strongly charged naphthalene and generated longer-range and larger jump-in and pull-off forces, but the charging extent was similar to contact charging (not shown in Figure \ref{fig:2}).

For polystyrene, to make sure that the charging is performed between identical materials, we used the polystyrene colloidal sphere to charge another polystyrene sphere, because we were unable to find a polystyrene flat surface with identical polymerization as the polystyrene spheres. We found that, similar to naphthalene, polystyrene spheres charge strongly between each other after both contact charging and tribocharging. The charge to mass ratio is calculated to be (4.6 $\pm$ 1.5)$\times$10$^{-6}$ C/kg for polystyrene (polystyrene sphere diameter 90 $\mu$m and density of sphere 1070 kg/m$^3$, Figure \ref{fig:5}), which is within the range reported by M\'endez-Harper et al., (2017).

\subsection{Charging between tholin}

\begin{figure}[h]
\centering
\includegraphics[width=30pc]{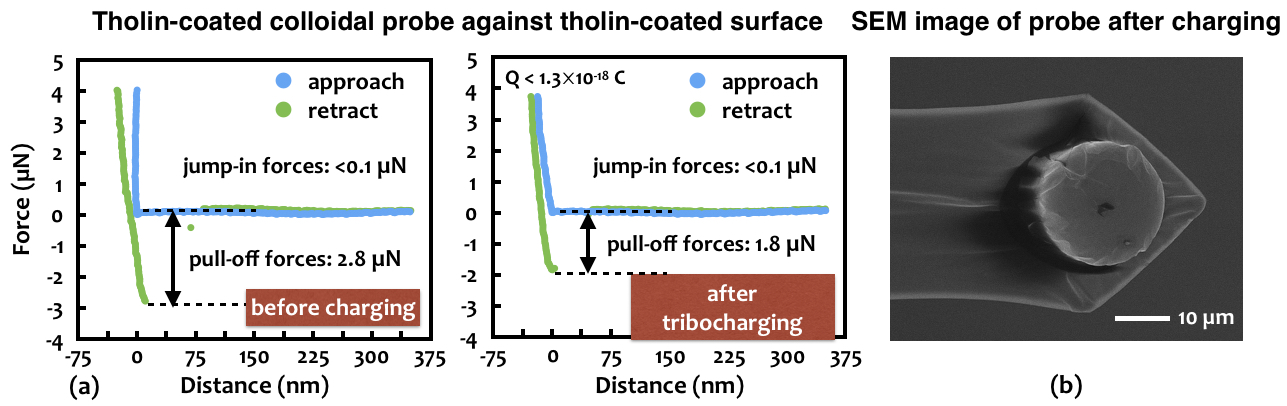}
\caption{(a) Force-distance curves taken between a tholin-coated colloidal probe and tholin-coated surface, before charging and after tribocharging. The jump-in forces are very small for both cases and the pull-off forces are smaller after tribocharging than before charging. (b) An SEM image of the tholin colloidal probe after multiple cycles of tribocharging (frictional rubbing), as shown in the image, visible cracks are developed indicating the brittle nature of tholin as measured by Yu et al., (2018).}
\label{fig:3}
\end{figure}

\begin{figure}[h]
\centering
\includegraphics[width=30pc]{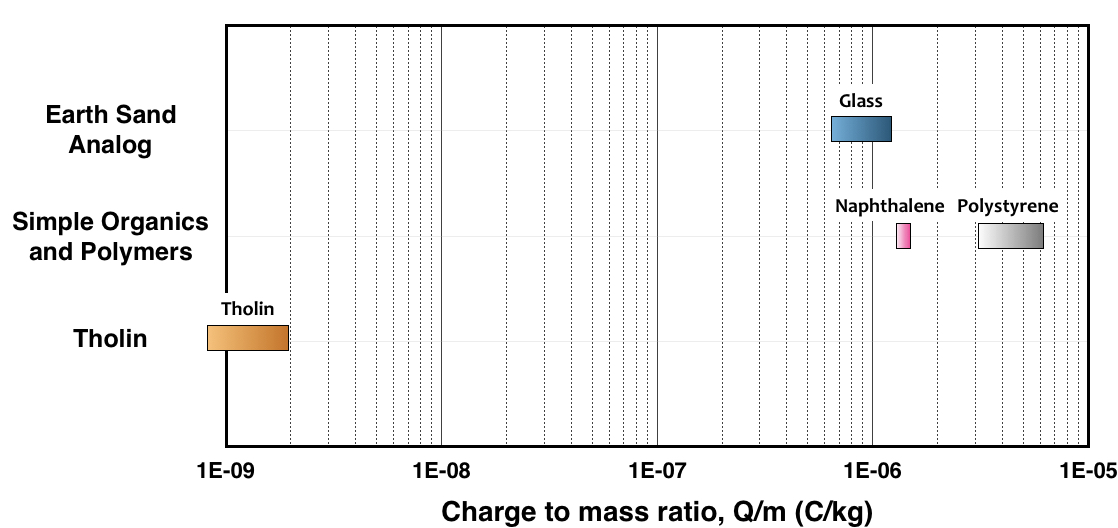}
\caption{Calculated charge to mass ratio for all measured materials. The color bar includes the standard deviation from the calculation. Since we did not detect any measurable electrostatic forces between tholin surfaces, the right edge of tholin's color bar gives the upper detection limit of the instrument.}
\label{fig:5}
\end{figure}

For the first time, we also performed charging experiments between Titan aerosol analog, tholin, which is a complex organic made of various chemical species (Cable et al., 2012). All the charging experiments and measurements were conducted in a dry nitrogen atmosphere (RH$<$1\%). Figure \ref{fig:3}(a) shows the set of force-distance curves taken between a coated tholin sphere and a smooth tholin coated surface, right before charging and after tribocharging. As shown in the force curves after charging, no long-range forces were observed. The jump-in forces are hardly detectable (less than 0.1 $\mu$N) both before charging and after tribocharging. Actually, we were unable to generate any long-range electrostatic attractive forces between tholin with either contact or tribocharging. Using the detection limit we derived in the methods section, we calculated the upper limit of the detected charge amount between tholin surfaces to be 1.3$\times$10$^{-18}$ C. The resulting charge to mass ratio is 1.9$\times$10$^{-9}$ C/kg (sphere diameter 80 $\mu$m and density of sphere 2500 kg/m$^3$, see Figure 5), which is two to three orders of magnitude smaller than the charge to mass ratio generated between glass, polystyrene, and naphthalene surfaces.

Note that in Figure \ref{fig:3}(a) the pull-off forces after charging (1.8 $\mu$N) are even smaller in magnitude than before charging (2.8 $\mu$N), which is likely caused by a difference in contact area or surface roughness after rubbing. The stochastic change of the magnitude of the pull-off forces does not directly relate to whether the long-range electrostatic forces are present or not. As shown in the SEM image of the tholin-coated colloidal probe after charging (Figure 4(b)), cracks are generated and the two contact surfaces could have different contact area before and after charging, leading to different pull-off forces. Electrostatic forces, however, are long-range in nature. If electrostatic forces are generated, a deviation of the force from the flat baseline will be detected as shown in Figure 2 and 3. But this was not detected between tholin surfaces. For tribocharging, we tried to increase the scan speeds (from 0.5 Hz to 10 Hz) and scan size (from 10 $\mu$m$\mathrm{\times}$10 $\mu$m to 50 $\mu$m$\mathrm{\times}$50 $\mu$m) to increase the contact area and rubbing frequency between the tholin surfaces, but we were still not able to generate any long-range electrostatic forces within the capacity of the instrument.

\section{Discussion}

\subsection{Charging Mechanism}
It was previously believed that contact charging and triobocharging only happened between different materials with distinct work functions (the energy required to remove an electron from the surface), which results in the ``triboelectric series" (Shaw, 1917). However, charging has been observed between spatially homogenous and chemically identical contacting insulators (Lowell and Rose-Innes, 1980; Lowell and Truscott, 1986; Komatsu et al., 2004; Apocada et al., 2010). However, the underlying driving mechanism for charge transfer between identical insulators is still unknown. Several proposed mechanisms include asymmetric rubbing (Lowell and Rose-Innes, 1980) and statistical variations in material properties (Apocada et al., 2010). The proposed transfer species include electrons, ions, and atomic, nanoscale, or microscale material bits (Lack and Mohan-Sankaran, 2011). However, on the local scale, it was found that the magnitude of charge on a charged surface may not be uniformly negative or positive; instead, the surface may consist of a mosaic of positively and negatively charged domains in nanoscale dimensions (Pollock et al., 1995; Baytekin et al., 2011). They observed that the charge density on each mosaic is relatively high, while the overall net charge is much weaker. They attributed the observation to the local chemical and nanomechanical inhomogeneity at or near the surface of the material (PDMS was used in the study). Tholin is a material with more complex chemical nature (it is made of a mixture of chemicals, see e.g., He et al., 2012; He \& Smith, 2014; Sebree et al., 2018), thus the mosaic pattern could be more complex for tholin compared to PDMS, leading to a weaker overall charge on the surface, because the mosaics cancel out each other. The instrument used here measured the overall charge between the surfaces rather than the charge domain distribution over the surfaces. For future studies we will attempt to measure these possible charge domains to confirm this hypothesis.

Our results suggest that simple organics like naphthalene and polystyrene and silicates like glass would charge more easily compared to tholin. Naphthalene and polystyrene would charge strongly by contact charging, while tholin does not charge under even the most violent abrasive rubbing and contacts that are allowed by our equipment. The electrical resistivity of all the test materials in this study are listed in Table 1. They are all insulators and their electrical resistivities are orders of magnitudes larger than conductors (e.g., metals have resistivities of 10$^{-8}$ $\Omega\cdot m$). So tholin should not dissipate charges faster than the other materials.


Although it is a powerful method for characterization of nanoscale interactions between single particles, our charging method with the AFM has a few limitations. First, the charging/rubbing rate (up to several hundred microns per second) is about two orders of magnitude slower than the natural charging processes in wind-blown sand (around several cm/s). The impact velocity of Titan sand is the about 0.5 m/s (Kok et al., 2012; J. Kok, personal communication), and the density of the grain is around 500-1400 kg/m$^3$ (Horst \& Tolbert, 2013; He et al., 2017). For AFM, the fastest rubbing gives the grain a rubbing velocity of 1000 $\mu$m/s (10 Hz scan rate with a 50 $\mu$m scan size). The size of the grain is also several times smaller than actual grains on Titan, thus the collisional energy of the AFM is at least 4 to 5 orders of magnitudes smaller than natural environments on Titan. So we are essentially measuring the lower limit of the triboelectric charges developed on tholin particles. It is likely that with higher collisional energies they could charge each other, but their charging capacity is definitely weaker than polystyrene, naphthalene, and even glass, because those materials already charge strongly with the low collisional energies that the AFM generated.

Second, even though we can study the interaction between two single grains, our method misses the multiple particle interactions in granular systems (Kok and Lacks, 2009), for which electrostatic charging is not only correlated with the bulk chemical properties of the materials, but also particle size distribution (Forward et al., 2009). We use spherical particles with size between 20 and 100 $\mu$m, which are smaller than the estimated optimum size range of sand grains on Titan (100-300 $\mu$m, Lorenz et al., 2006). But the sphere to flat charging system can simulate a small grain charging a bigger grain (flat surface corresponds to a grain of infinite radius) in a granular flow system. Additionally, other materials with similar grain sizes tribocharge easily (naphthalene, polystyrene, and glass), while tholin does not. This indicates that as a material, tholin has weaker tribocharging capacity compared to common polymers, simple organics, and Earth silicate sand.

Third, the AFM can only characterize the magnitude of the net electrostatic forces between the two surfaces, without identifying the orientation of the charge domains on the two surfaces or the detailed surface charge distribution. More advanced characterization methods such as Electrostatic Force Microscopy (EFM) and Kelvin Probe Force Microscopy (KPFM) can map the charge distribution on the material surface and provide more information on charge domains, which may be useful for future investigations.

\subsection{Implications for aeolian processes on Titan}
Dunes on Titan are spectrally dark (Soderblom et al., 2007), while most simple organics (e.g., solid benzene, ethylene, acetylene) are spectrally bright (Clark et al., 2010), thus they are unlikely to be the only constituents of Titan sand. A spectrally darker material like tholin would be a better analog for Titan's surface sand than the spectrally bright simple organics. If that is the case, our study would suggest that Titan sand may be unlikely to be formed by triboelectric coagulation from the aerosol particles, while high cohesion between the material would instead be more viable (Yu et al., 2017b). As discussed in M\'endez-Harper et al., (2017), Titan's cold temperature and low water humidity could help preserve the charges better if electrostatic charges are produced. However, our experiments show that under room temperature and low water humidity, tholin has minimal tribocharging capacity (Figure 5). Under the same conditions, glass, an analog for Earth sand, generates detectable electrostatic charges. Thus, tribocharging of Titan sand may not be as strong as previously expected by M\'endez-Harper et al., (2017). The triboelectric effect could be also weaker for Titan sand compared to silicate Earth sand. In other words, tribocharging may not significantly increase the threshold wind speed on Titan compared to on Earth.

\subsection{Implications for lightning on Titan}
Electrification of granular materials can result in the build-up of significant electrical charges on a material surface. When the charge becomes high enough, the electrical field that develops between surfaces can exceed the threshold dielectric strength, leading to electrical breakdown and discharge of the gas in between. Convective clouds are common sources for lightning, but sand storms could also generated lightning if the breakdown voltage is reached through triboelectric charging (Tokano et al., 2001). The dielectric strength of Titan's atmosphere in the troposphere is similar to Earth's troposphere (Fisher et al., 2004), around 3$\times$10$^6$ V/m. However, no lightning has been detected on Titan so far (Fischer and Gurnett, 2011), even though global sand storms were observed during Titan's equinox (Rodriguez et al., 2018). Lightning often occurs in very energetic dusty systems, while our experimental system cannot reproduce similar energetic environments. However, lightning is also linked to a material's tribocharging capacity. Our results suggest that Titan's sand may not be strongly charged due to the low charging capacity of tholin, which may contribute to the observation that no lightning has been detected so far on Titan. Thus electrical discharge generated by sand storms may not be an issue for future surface or near-surface missions on Titan through tribocharging.

 
\section{Conclusion}
To understand the effect of electrification for dune formation on Titan, we performed contact and triboelectric charging between possible organic candidates using colloidal atomic force microscopy. This technique can characterize individual charging processes and avoid contamination between the surfaces. The simple organic naphthalene and the simple hydrocarbon polymer polystyrene were found to be charged very easily against themselves from repeated contacts, so is for Earth silicate sand. While the complex organic, tholin, did not generate any detectable electrostatic attraction ($<$10$^{-15}$ C) after even the most violent tribocharging that the instrument is capable of. Our study indicates that the charging abilities of Titan sand may be much weaker than previously expected from M\'endez-Harper et al., (2017). If Titan sand is similar in composition to the complex organic, tholin, then the triboelectric effect could be weaker on Titan than on Earth. Thus, triboelectric charging and the resulting electrostatic forces may not be as important as van der Waals forces (Yu et al., 2017a,b) for Titan sand. Thus the threshold wind speed will not increase due to sand tribocharging on Titan. Our study also agrees with the observation that no lightning was generated on Titan (Fischer and Gurnett, 2011), even though large sand storms were observed during the search period (Rodriguez et al., 2018).

\section{Acknowledgments}
X. Yu is supported by the 51 Pegasi b Fellowship from the Heising-Simons Foundation. P. McGuiggan is supported by the 3M Nontenured Faculty Grant. We would like to thank Jason J. Benkoski for his valuable comments. We continue to be thankful for Nathan Bridges, who started us on the journey to the fabulous dune worlds.

\end{document}